 \documentclass[preprint]{aastex}
\usepackage[colorlinks, linkcolor=blue, citecolor=blue, breaklinks=true]{hyperref}
 \usepackage{ulem}
\usepackage{xcolor}

\definecolor{gg}{gray}{0.5}

\shorttitle{Variations in solar wind fractionation as seen by ACE/SWICS and the implications for Genesis Mission results}
\shortauthors{Pilleri et al.}

\begin{document}


\title{Variations in solar wind fractionation as seen by ACE/SWICS over a solar cycle and the implications for Genesis Mission results}


\author{P. Pilleri\altaffilmark{1}}
\affil{Los Alamos National Laboratory, P.O. Box 1663, Los Alamos (NM) 87545, USA}

\author{D. B. Reisenfeld}
\affil{University of Montana,
  Department of Physics \& Astronomy,
  32 Campus Drive,
  Missoula, Montana, USA}

\author{T. H. Zurbuchen, S. T.  Lepri, P. Shearer,  J. A. Gilbert}
\affil{Department of Atmospheric, Oceanic, and Space Sciences, 2455 Hayward St., 
University of Michigan, Ann Arbor, MI 48109, USA}

\author{R. von Steiger}
\affil{International Space Science Institute, Hallerstrasse 6, CH-3012 Bern, Switzerland}
\affil{Physikalisches Institut, University of Bern, Sidlerstrasse 5, CH-3012 Bern, Switzerland}

\author{R. C. Wiens}
\affil{Los Alamos National Laboratory, P.O. Box 1663, Los Alamos (NM) 87545, USA}


\altaffiltext{1}{Current address: Universit\'e de Toulouse; UPS-OMP; IRAP; 9 Av. colonel Roche, BP 44346, F-31028 Toulouse cedex 4, France. email: \url{paolo.pilleri@irap.omp.eu}.}


\begin{abstract}
We use ACE/SWICS elemental composition data to compare the variations in solar wind fractionation as measured by SWICS during the last solar maximum (1999-2001), the solar minimum (2006-2009) and the period in which the Genesis spacecraft was collecting solar wind (late 2001 - early 2004). We differentiate our analysis in terms of  solar wind regimes (i.e.  originating from interstream or coronal hole flows, or coronal mass ejecta).  Abundances are normalized to the low-FIP ion magnesium to uncover correlations that are not apparent when normalizing to high-FIP ions.

We find that relative to magnesium, the other low-FIP elements are measurably fractionated, but the degree of fractionation does not vary significantly over the solar cycle.  For the high-FIP ions, variation in fractionation over the solar cycle is significant: greatest for Ne/Mg and C/Mg, less so for O/Mg, and the least for He/Mg.  When abundance ratios are examined as a function of solar wind speed, we find a strong correlation, with the remarkable observation that the degree of fractionation follows a mass-dependent trend.   We discuss the implications for correcting the Genesis sample return results to photospheric abundances.
\end{abstract}


\keywords{(Sun:) solar wind -- Sun: abundances}



\section{Introduction}

The elemental abundances in the outer convective zone of the Sun are significantly different compared to  those observed in the solar wind (SW). Phenomenologically,
the variation of the elemental abundance in the solar wind ($X_{sw}$) relative to its photospheric value ($X_{phot}$),  {\it fractionation}, appears to be connected with the element first ionization potential \citep[FIP,][]{meyer93}. In particular, elements with FIP above 10\,eV (the so-called high-FIP elements) are depleted relative to low-FIP ones. Because of its ease of detection and hence low uncertainty, oxygen has been often used as a normalizing element 
to demonstrate {\it enrichments} or {\it depletion} in the SW compared to the photosphere 
\citep[see, for instance,][]{meyer85, vonsteiger00}. However,  to study the relative fractionation of elements and interpret them in terms of physical processes, low-FIP elements such as Mg or Fe are more suitable for normalization.

The FIP effect is attributed to the preferential ionization of low-FIP elements in the chromosphere, which are then entrained into the solar wind via 
mechanisms such as ambipolar diffusion \citep[\textit{e.g.},][]{marsch95}, inefficient Coulomb drag \citep{buergi86, buergi92, bodmer98, bodmer00} or wave-particle interaction \citep{schwadron99, laming04}. 
The SW elemental fractionation depends on the region of the Sun from which the flow originates.  For the quasi-stationary wind, interstream (IS) flow, originating near closed-loop boundaries and manifesting as a slow wind ($v_p < 500$ km/s is a typical cutoff) shows greater fractionation than coronal hole (CH) flow, which manifests as a fast wind ($v_p > 500$ km/s) \citep[][{and references therein}]{zurbuchen07}.   In fact, some argue that the very fast solar wind ($v_p \sim 700$ km/s) associated with polar coronal holes is essentially unfractionated compared to the photosphere \citep{gloeckler07}.  The transient flow associated with coronal mass ejections (CMEs) often exhibits greater fractionation than IS flow \citep{reisenfeld03, richardson04}.  In the ecliptic, there is also significant temporal variability of the SW elemental composition, which may be cyclic with the phase of the solar cycle \citep{vonsteiger00, lepri13, shearer14}.

Finding a theoretical model that consistently and quantifiably explains the fractionation observations has proven challenging. The wave-particle model of    \citet{laming04,laming09} shows some promise in reproducing specific elemental abundances in the different wind types.  Although generally reproducing the observed FIP trends, the species-specific agreement is still only fair, and the model has as a free parameter the (unknown) magnetic wave energy spectrum present in the corona.  

The need for an accurate fractionation theory has become all the more pressing because of the {\it Genesis} mission, which collected a sample of the solar wind with the goal of deriving from it the solar composition \citep{burnett03}.  The primary objective of the mission is to determine the isotopic compositions of oxygen and nitrogen in the primordial solar nebula.  Since the Sun contains over $99\%$ of the mass of the solar system, the photosphere is considered an excellent proxy for the primordial solar nebula.   A secondary objective is to make accurate measurements of the solar elemental abundances.  Genesis collected solar material at the L1 point between the Earth and Sun for 2.3 years between November 30, 2001 and April 1, 2004, when the solar cycle was at and declining from the solar maximum of solar cycle 23 \citep{reisenfeld07}. The Genesis sample constitutes the largest fluence of solar wind material collected to date for determining SW isotopic and elemental composition and fractionation.   Elemental abundances have been measured with a typical accuracy of $\sim 5\%$ \citep{heber14}, and isotopic ratios to an accuracy of better than $1\%$ \citep{heber09}.  

When Genesis was proposed, It was understood that it would be necessary to correct the SW elemental composition to photospheric values.  To that end, in addition to collectors exposed during the entire mission (the 'bulk' collectors), certain collectors were exposed only during one of three specific flow regimes: IS, CH, or CME, with the thought to use the measured differences as an aide in resolving the fractionation problem.  Also, of the three regimes, it was hoped that the CH sample would be the least fractionated relative to the photosphere.  Even if the CH sample is fractionated, since coronal holes have a relatively simple magnetic structure, then possibly this sample would be the most easily corrected to photospheric values.

Prior to Genesis, there was only sparse evidence that the SW isotopes were fractionated relative to the photosphere \citep{kallenbach97, kallenbach98}.  The regime-specific samples would then be a check on whether there was isotopic fractionation.  Even if there were fractionation, it was expected that the CH sample would be the least fractionated, if at all.  Recent analysis of the regime-specific samples for noble gas isotopes now show a small but highly statistically significant ($ >~3 \sigma$) difference in isotopic abundances between regimes \citep{heber12}, and therefore between the SW and photosphere.  At the present time, there is no theory that can quantitatively model isotopic fractionation.

The purpose of this article is to gain further insight into the nature of solar wind elemental fractionation and its temporal variability, with particular focus on comparison of the Genesis collection period to other phases of the solar cycle.  Two aspects of elemental composition will be investigated.  One is the variation of composition with phase of the solar cycle.  These changes in the SW composition over the solar cycle raise the question of whether the Genesis sample is representative of the whole solar cycle (and thus require a single correction factor per element/isotope), or if a solar cycle phase-dependent correction factor is needed.  The second aspect of composition we wish to consider is variation with solar wind speed.  Although it is common to separate the compositional properties of the quasi-stationary SW into the two discrete categories of ÔfastÕ and ÔslowÕ, it has been shown that this description may be too simplified. Rather, the solar wind composition varies as a function of speed, both in terms of charge state \citep{gloeckler03} and elemental composition \citep{reisenfeld07}.  Here, we take a more systematic look at the relationship between SW speed and elemental composition.  We normalize the elemental abundances to magnesium, a low-FIP element, to investigate the differences of abundances among low-FIP elements  and how they vary with time and with solar wind speed.   To achieve this, we use data from the  Solar Wind Ion Composition Spectrometer (SWICS, \citep{gloeckler98}) onboard the {\it Advanced Composition Explorer} \citep[ACE,][]{stone98} which provides us with a continuous set of in situ measurements since its launch in late 1997.  

The article is organized as follows: section 2 describes the SWICS data sample and our method of analysis. Section 3 describes the results, comparing the FIP fractionation for the bulk solar wind and the different SW regimes (CH, IS and CME) during the Genesis collection period with the solar minimum and solar maximum of cycle 23. Section 4 provides a brief discussion of the results and the impact in the analysis of the SW fractionation theories. Finally, Section 5 presents the conclusions.

\section{Data sample and analysis}

The elemental composition measurements presented here are obtained from the SWICS instrument on board ACE.  
We use the recently released SWICS data set that includes important   improvements  \citep{shearer14} over previous versions. Most importantly, the new data includes improvements of the statistical methodology, which has important impact especially in low-abundance ions which have non-Gaussian errors because of  low count-rates. These data are available at the ACE Science Data Center (\url{http://www.srl.caltech.edu/ACE/ASC/}). Typically, the publicly available dataset provides  elemental abundances relative to oxygen for both low FIP (Fe, Si and Mg), intermediate FIP (C), and high-FIP elements (Ne, Ne). 

The data revision is significant in that prior conclusions \citep[e.g. those in][]{lepri13} need to be reconfirmed.  A thorough review of papers done with data prior to the \citep{shearer14} methodology \citep[such as][]{lepri13} are currently underway.

\subsection{Regime Selection}
\label{sec_reg_select}
\begin{figure*}[!t]
 \centering
 \noindent\includegraphics[width=\textwidth]{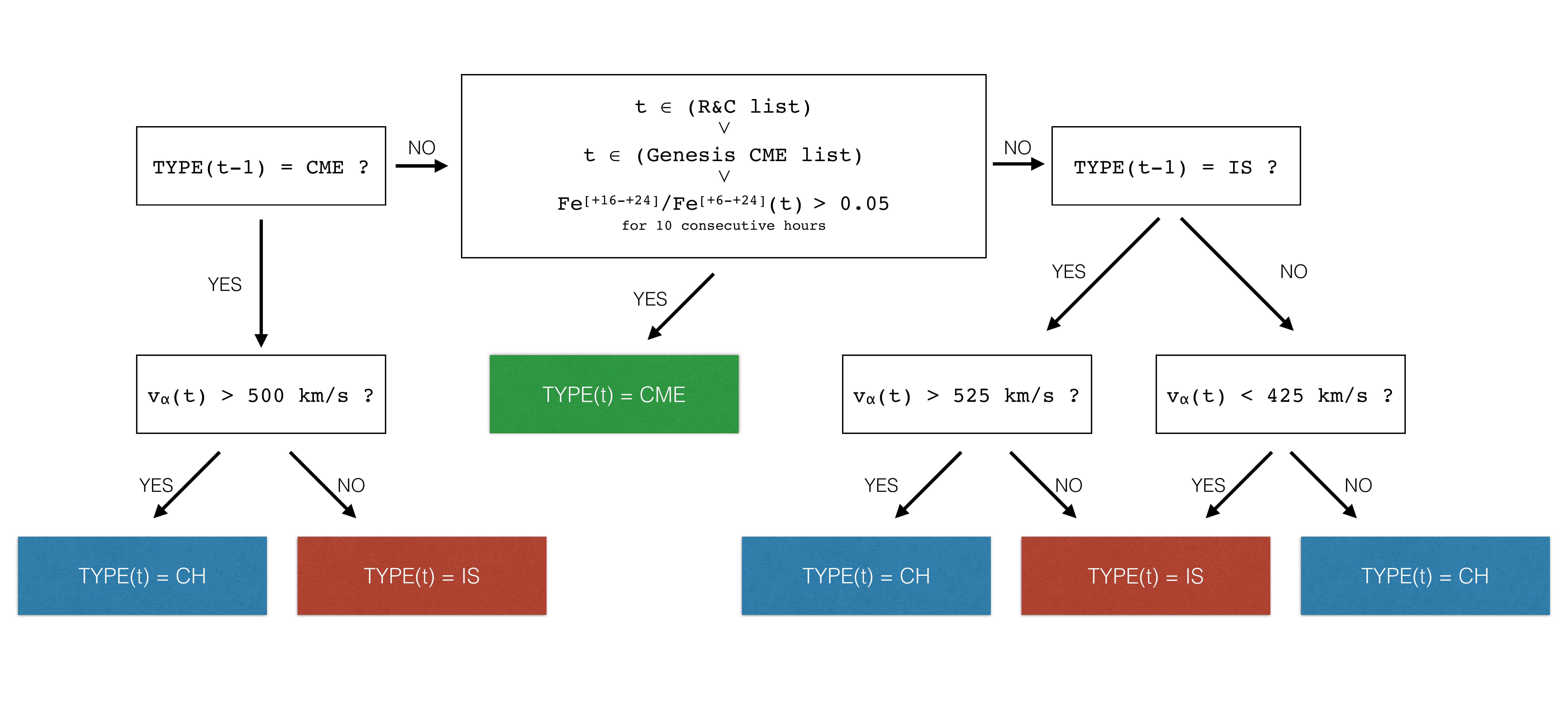}
 \caption{Modified genesis algorithm used to separate the different flows of SW.}
\label{fig_genalg}
 \end{figure*}

To explore how the solar wind composition varies over the course of the solar cycle, we will focus on three time periods within the ACE data collection period: solar maximum (January 1, 1999 Ð December 31, 2000),  the Genesis collection period (Nov 30, 2001 Ð April 1, 2004), and solar minimum (January 1, 2007 Ð December 31, 2009).  Since our ultimate objective is to interpret solar wind fluence measurements taken from the Genesis regime collectors, we sort the ACE/SWICS composition measurements for the periods of interest using as close to the same algorithm as was used for the Genesis mission during its flight as described in \citet{neugebauer03} and schematized in Fig.\,\ref{fig_genalg}. This method analyzes the temporal variation of the SW speed to determine transitions  between fast (CH) and slow (IS) solar wind, by taking into account its hydrodynamical evolution along a stream interface. At a slow-to-fast transition, the speed threshold to discriminate between IS and CH wind is set relatively high (525 km/s) to take into account the fact that the fast wind has accelerated the slow stream. Similarly, in the fast-to-slow transition, the threshold is set low (425 km/s) to take into account the deceleration of fast wind due to the rarefaction that forms between the fast wind and the following slow wind.  

\citet{reisenfeld13} shows that this "hysteresis" algorithm did a superior job of sorting the solar wind between solar source regions than just simple categorization of regimes by a speed threshold, as done in other studies \citep[e.g.][]{tu95, lepri13}.  Other methods attribute SW flow to CH and IS origin by including the analysis of the SW  composition \citep[in particular, the O$^{7+}$/O$^{6+}$ ratio,][]{zurbuchen02, zhao09}.   Although these criteria may be more accurate than the Genesis algorithm, we are specifically addressing the relevance of the Genesis data, and therefore we apply the algorithm that was used by that mission. 

The Genesis criterion for detecting possible plasma associated with CMEs  was based on a weighting function that  depended on the presence of a subset of the following three criteria: a low proton temperature, a high alpha/proton ratio and the presence of bi-directional electron flow. Because the latter is not a readily available data product from ACE, we labeled as possible CMEs all the data points that  satisfy one of the following criteria: $i)$ data taken within the specific time intervals from the Richardson-Cane list of CMEs \citet{richardson10}; $ii)$ data where the density ratio Fe$^{[+16 -  +24]}$/Fe$^{[+6 - +24]}$ is greater than 0.05 for at least 10 hours (5 consecutive 2-hour data points). This criterion is a version of that used in \citet{lepri01}, and more recently in \citet{shearer14}, although with slightly different thresholds  and time interval; and $iii)$  data taken at times during the actual Genesis mission when the Genesis onboard algorithm catalogued the flow as a CME.  This latter criterion makes use of the Genesis bi-directional streaming measurements, which are not available in the SWICS data. Comparing the three criteria for the time interval that Genesis was active, we find that the actual Genesis algorithm (criterion $iii$) detects the largest number of CMEs, followed by the analysis of the Fe charge state distribution (criterion $ii$) and finally by the RC list  \citep[see also][]{reisenfeld13}. To be as conservative as possible in the analysis of IS and CH flows, any data point that satisfy at least one criterion is marked as CME.

\begin{table*}
\renewcommand{\arraystretch}{1.65}
\caption{First ionization potentials and photospheric abundances normalized to magnesium \citep[derived from][]{asplund09}.  Percent uncertainty in the quantity [X/Mg]$^*$ includes the {\it ACE}/SWICS 15\% instrumental uncertainty in quadrature.}
\label{tab_phot_stats}
\begin{center}
\begin{tabular}{c  c  c  c  c}
\hline \hline
Element & FIP (eV) & Phot. Abundance$^1$ & \% Uncert& \% Uncert in [X/Mg]$^*$\\
\hline
Mg	&	7.65	&	1.00		&	--	&	--\\
Fe	&	7.87	&	0.794	&	13	&	25\\
Si	&	8.15	&	0.813	&	12	&	24\\
C	&	11.3	&	6.76		&	15	&	26\\	
O	&	13.6	&	12.3		&	15	&	21\\
Ne	&	21.6	&	2.14$^2$	&	25	&	33\\
He	&	24.6	&	2140		&	10	&	23\\
\hline
\end{tabular}
\end{center}
 $^1$ Photospheric abundances normalized to Mg.\\
$^2$ indirect photospheric estimate based on coronal lines.
\end{table*}

\begin{table*}
\renewcommand{\arraystretch}{1.65}\addtolength{\tabcolsep}{1.5pt}
\caption{Solar wind speed and elemental abundance ratios, as a function of time period and solar wind regime.}
\label{tab_period_stats}
\begin{center}
\begin{tabular}{rcccc|cccc|cccc}
\hline \hline
		& \multicolumn{4}{c}{Solar Max} & \multicolumn{4}{c}{Genesis} & \multicolumn{4}{c}{Solar Min} \\ 
\cline{2-5} \cline{6-9} \cline{10-13}
     Parameter       &      Bulk  &         IS  &         CH  &        CME  &       Bulk  &         IS  &         CH  &        CME  &       Bulk  &         IS  &         CH  &        CME  \\
\hline
    $<v_{\alpha}>$&              459  &        392  &        562  &        458  &        511  &        402  &        598  &        496  &        426  &        361  &        563  &        408\\
  $\sigma_v$&       106  &         50  &         86  &        107  &        134  &         54  &        105  &        140  &        113  &         53  &         79  &        102\\
 \hline
  Mean(Fe/Mg)&      1.16  &       1.27  &       1.10  &       1.08  &       1.23  &       1.41  &       1.18  &       1.17  &       1.20  &       1.24  &       1.14  &       0.76\\
 Lower(Fe/Mg)&      0.77  &       0.86  &       0.80  &       0.69  &       0.83  &       0.95  &       0.88  &       0.74  &       0.78  &       0.77  &       0.87  &       0.35\\
 Upper(Fe/Mg)&      1.74  &       1.88  &       1.53  &       1.71  &       1.82  &       2.08  &       1.59  &       1.85  &       1.85  &       2.01  &       1.49  &       1.65\\
\hline
  Mean(Si/Mg)&      1.29  &       1.32  &       1.38  &       1.20  &       1.41  &       1.47  &       1.49  &       1.31  &       1.32  &       1.27  &       1.45  &       1.10\\
 Lower(Fe/Mg)&      1.01  &       1.06  &       1.10  &       0.93  &       1.11  &       1.18  &       1.21  &       1.01  &       0.96  &       0.91  &       1.11  &       0.67\\
 Upper(Fe/Mg)&      1.64  &       1.65  &       1.72  &       1.54  &       1.80  &       1.83  &       1.84  &       1.70  &       1.83  &       1.78  &       1.87  &       1.78\\
\hline
   Mean(C/Mg)&      0.49  &       0.48  &       0.59  &       0.43  &       0.56  &       0.53  &       0.71  &       0.47  &       0.70  &       0.64  &       0.85  &       0.55\\
 Lower(Fe/Mg)&      0.31  &       0.33  &       0.41  &       0.25  &       0.35  &       0.37  &       0.51  &       0.26  &       0.50  &       0.47  &       0.66  &       0.34\\
 Upper(Fe/Mg)&      0.77  &       0.70  &       0.86  &       0.74  &       0.91  &       0.75  &       0.97  &       0.83  &       0.96  &       0.87  &       1.09  &       0.90\\
\hline
   Mean(O/Mg)&      0.47  &       0.47  &       0.53  &       0.42  &       0.54  &       0.51  &       0.62  &       0.47  &       0.61  &       0.56  &       0.73  &       0.47\\
 Lower(Fe/Mg)&      0.34  &       0.37  &       0.38  &       0.29  &       0.38  &       0.40  &       0.46  &       0.33  &       0.45  &       0.41  &       0.59  &       0.29\\
 Upper(Fe/Mg)&      0.65  &       0.61  &       0.74  &       0.61  &       0.75  &       0.65  &       0.84  &       0.69  &       0.83  &       0.76  &       0.91  &       0.75\\
\hline
  Mean(Ne/Mg)&      0.32  &       0.31  &       0.30  &       0.34  &       0.35  &       0.32  &       0.37  &       0.37  &       0.49  &       0.50  &       0.49  &       0.47\\
 Lower(Fe/Mg)&      0.23  &       0.23  &       0.21  &       0.25  &       0.25  &       0.24  &       0.25  &       0.27  &       0.36  &       0.36  &       0.36  &       0.30\\
 Upper(Fe/Mg)&      0.44  &       0.41  &       0.44  &       0.47  &       0.50  &       0.43  &       0.53  &       0.50  &       0.69  &       0.70  &       0.66  &       0.75\\
\hline
  Mean(He/Mg)&      0.26  &       0.26  &       0.27  &       0.26  &       0.28  &       0.28  &       0.30  &       0.25  &       0.23  &       0.19  &       0.33  &       0.14\\
 Lower(Fe/Mg)&      0.16  &       0.17  &       0.18  &       0.15  &       0.17  &       0.18  &       0.21  &       0.14  &       0.12  &       0.11  &       0.23  &       0.06\\
 Upper(Fe/Mg)&      0.43  &       0.41  &       0.41  &       0.46  &       0.45  &       0.43  &       0.43  &       0.46  &       0.41  &       0.35  &       0.48  &       0.34\\
\hline
        f$_{FIP}$&              2.62  &       2.67  &       2.33  &       2.80  &       2.44  &       2.65  &       2.03  &       2.72  &       1.96  &       2.12  &       1.63  &       2.08\\
  \end{tabular}
\end{center}
\end{table*}

\begin{table*}
\renewcommand{\arraystretch}{1.65}\addtolength{\tabcolsep}{3pt}
\caption{Percent differences of the mean abundance ratios between periods.   Differences in  bold are  are greater than the instrumental uncertainty of 15\% for O/Mg and 21\% for the other elements.  The differences below instrumental uncertainty are shown in light grey to emphasise their non significance.}
\label{tab_frac_diff}
\begin{center}
\begin{tabular}{lcccc|cccc|cccc}
\hline \hline
		&\multicolumn{4}{c}{Genesis vs. Max}&\multicolumn{4}{c}{Genesis vs. Min}&\multicolumn{4}{c}{Max vs. Min}\\
\cline{2-5}\cline{6-9}\cline{10-13}
		& Bulk & IS & CH & CME & Bulk & IS & CH & CME  & Bulk & IS & CH & CME\\
\hline
(Fe/Mg)$^*$	& \color{gg}{6} &  \color{gg}{-11} &  \color{gg}{7} &  \color{gg}{8} &  \color{gg}{2} &  \color{gg}{-9} &  \color{gg}{3} & {\bf 35} & \color{gg}{-3} &  \color{gg}{2} &  \color{gg}{-4} & {\bf 30} \\
(Si/Mg)$^*$	&  \color{gg}{9} &  \color{gg}{10} &  \color{gg}{7} &  \color{gg}{8} &  \color{gg}{6} &  \color{gg}{14} &  \color{gg}{3} &  \color{gg}{16} &  \color{gg}{-2} & \color{gg}{4} &  \color{gg}{-5} &  \color{gg}{8} \\
(C/Mg)$^*$	&  \color{gg}{13} &  \color{gg}{9} &  \color{gg}{17} & 3 & {\bf -25} & {\bf -21} &  \color{gg}{-20} &  \color{gg}{-17} & {\bf -43} & {\bf -33} & {\bf -44} & {\bf -28}\\
(O/Mg)$^*$	&  \color{gg}{13} &  \color{gg}{8} &  \color{gg}{15} & \color{gg}{11} &  \color{gg}{-13} &  \color{gg}{-10} & {\bf -18} &  \color{gg}{0} &  {\bf -30} & {\bf -19} & {\bf -38} &  \color{gg}{-12} \\
(Ne/Mg)$^*$	&  \color{gg}{9} &  \color{gg}{3} &  \color{gg}{19} &  \color{gg}{8} & {\bf -40} & {\bf -56} & {\bf -32} & {\bf -27} & {\bf -53} & {\bf -61} & {\bf -63} & {\bf -38}\\
(He/Mg)$^*$	&  \color{gg}{7} &  \color{gg}{7} &  \color{gg}{10} &  \color{gg}{-4} &  \color{gg}{18} & {\bf 32} &  \color{gg}{-10} & {\bf 44} &  \color{gg}{12} & {\bf 27} & {\bf -22} & {\bf 46}\\
\hline
\end{tabular}
\end{center}
\end{table*}

\begin{table*}
\renewcommand{\arraystretch}{1.65}\addtolength{\tabcolsep}{3pt}
\caption{Average of the absolute percent differences of the {\it speed-specific} mean abundance ratios between periods, $\Delta X$.  The speed bins are shown in Figure~\ref{fig_histograms}.  Note these quantities are all positive, since absolute differences are taken, unlike the values in Table~\ref{tab_frac_diff}. Differences in {\bf bold} are greater than the instrumental uncertainty of 15\% for O/Mg and 21\% for the other elements, whereas the non significant variations are shown in light grey.  There is no comparison to solar minimum CMEs because there were not enough CMEs during solar minimum for meaningful statistics within individual speed bins.}
\label{tab_frac_diff_speed}
\begin{center}
\begin{tabular}{lcccc|ccc|ccc}
\hline \hline
		&\multicolumn{4}{c}{Genesis vs. Max}&\multicolumn{3}{c}{Genesis vs. Min}&\multicolumn{3}{c}{Max vs. Min}\\
\cline{2-5}\cline{6-8}\cline{9-11}
		& Bulk & IS & CH & CME & Bulk & IS & CH & Bulk & IS & CH\\
\hline
(Fe/Mg)$^*$	&  \color{gg}{9} &  \color{gg}{10} &  \color{gg}{10}&  \color{gg}{10} &  \color{gg}{7} &  \color{gg}{13} & \color{gg}{6} &  \color{gg}{6} &  \color{gg}{5} &  \color{gg}{7}\\
(Si/Mg)$^*$	&  \color{gg}{10} &  \color{gg}{10} & \color{gg}{9} & \color{gg}\color{gg}{9} &  \color{gg}{5} &  \color{gg}{10} &  \color{gg}{4} &  \color{gg}{9} &  \color{gg}{3} &  \color{gg}{9}\\
(C/Mg)$^*$	&  \color{gg}{13} &  \color{gg}{5} &  \color{gg}{15} &  \color{gg}{13} & {\bf 40} & {\bf 35} & {\bf 23} & {\bf 52} & {\bf 41} & {\bf 38} \\
(O/Mg)$^*$	&  \color{gg}{13} &  \color{gg}{7} &  \color{gg}{14} &  \color{gg}{13} & {\bf 24} & {\bf 17} & {\bf 19} & {\bf 40} & {\bf 25} & {\bf 34} \\
(Ne/Mg)$^*$	&  \color{gg}{11} &  \color{gg}{3} &  \color{gg}{14} &  \color{gg}{9} & {\bf 39} & {\bf 54} & {\bf 35} & {\bf 52} & {\bf 57} & {\bf 55} \\
(He/Mg)$^*$	&  \color{gg}{11} &  \color{gg}{6} &  \color{gg}{11} & \color{gg}{14} & \color{gg}{17} &  \color{gg}{13} &  \color{gg}{12} & {\bf 23} &  \color{gg}{13} & {\bf 21} \\
\hline
\end{tabular}
\end{center}
\end{table*}

\section{Analysis and discussion}

In this section, we examine SW fractionation as a function of time and velocity for the more abundant elements observed by ACE/SWICS.   We consider three low-FIP  (Mg, Fe, Si), two intermediate FIP (O, C) and two high-FIP (Ne, He) elements.

When dealing with SW fractionation effects, it is customary to reference the measured abundance of the elements to their abundance in the photosphere: 

$$ X^* = \frac{X_{sw}}{X_{phot}},$$

where $X_{sw}$ and $X_{phot}$ refer to the abundance of a given element, normalized to a reference element.  Here, we normalize the elemental abundances to magnesium, the lowest FIP element in the ACE/SWICS dataset.  This is in order to emphasize how abundance differences among the low-FIP elements vary with time and with speed, thus emphasizing differences in fractionation due to effects beyond FIP.  The three low-FIP elements considered here, Mg, Fe and Si, have ionization potentials spanning a range of just 0.50 eV: 7.65 eV, 7.90 eV and 8.15 eV, respectively, so particularly for these elements, fractionation differences, if present, should originate from a different source. Furthermore, while oxygen is often used for normalization, as it is an intermediate FIP element, abundances normalized in this way do not show the low- to high-FIP fractionation as clearly as if they are normalized to a low-FIP element.

Table~\ref{tab_phot_stats} gives the first ionization potentials and photospheric abundances (and uncertainties) relative to Mg \citep{asplund09} for the elements considered in this study.  Table~\ref{tab_phot_stats} also gives the absolute uncertainties  (1$\sigma$) in the fractionation ratios $X^*$, derived by combining the SWICS instrument uncertainty and the photospheric uncertainties reported by \citet{asplund09}.  In what follows, discussion of the variation in abundances compared to the {\it photosphere} should be measured against the {\it absolute} uncertainties in $X^*$; otherwise discussion of differences in $X^*$ should be measured against the SWICS instrument uncertainty, which for the elements considered here is assumed to be 15\%. In general, these error bars are both dependent on counting statistics - which are also found on the ACE data center, as well as systematic errors, which are species dependent \citep[for details, see][]{vonsteiger11}. This uncertainty refers to the $X$/O ratio data that is retrieved from the archive. In the following, we normalize to magnesium, which is calculated from the ACE ratios: ($X$/O) / (Mg/O).   In this case, the significance level is 21\% (15\%$ \times \sqrt{2}$) using standard error propagation (except, of course when $X$ = O, in which case the uncertainty is still 15\%).  Note the statistical uncertainties are negligible compared to the systematic uncertainties due to the large size of the data sample considered in this study.

\subsection{Solar wind fractionation over time}

\begin{figure*}[!h]
 \centering
 \noindent\includegraphics[trim = 2cm 1cm 2cm 3cm, width=0.8\textwidth]{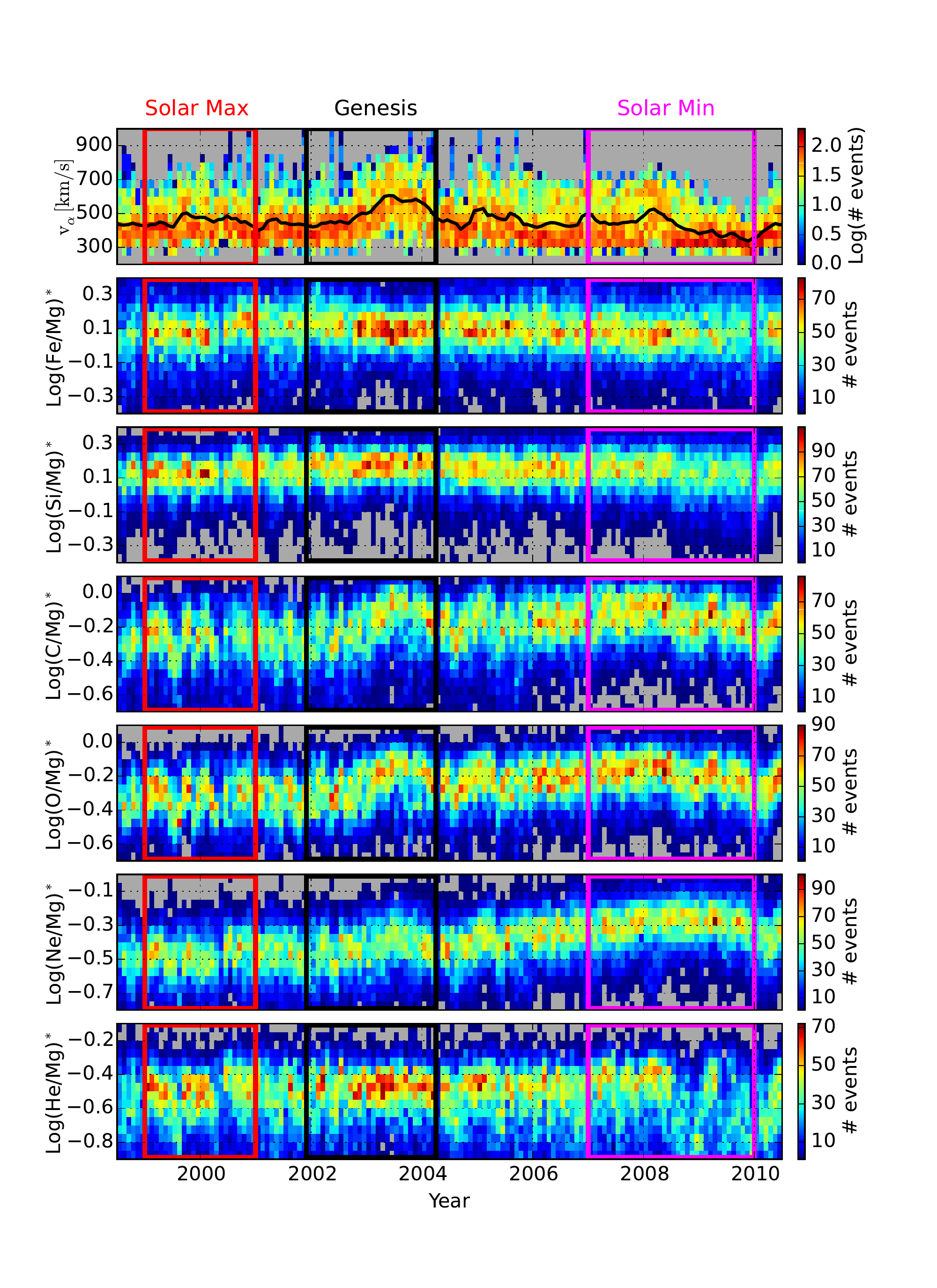}
 \caption{Top Panel: Logarithmic density of observations (i.e., the number of He$^2+$ measurements) of solar wind alpha speed on monthly intervals from 1999 to 2010.5. Following Panels: Density of observations of abundance ratios on monthly intervals. The "$\#$ of events" scale  indicates the number of simultaneous valid measurements of Mg and element $X$).  The elements are placed in order (top-to-bottom) of increasing FIP.}
\label{fig_time_variations_rat}
 \end{figure*}

We first examine the variation of the SW fractionation ratios relative to Mg in the three time intervals defined previously: solar maximum, the Genesis collection period, and solar minimum.  Fig. \ref{fig_time_variations_rat} shows the monthly distribution of fractionation ratios  (Fe/Mg)$^*$, (Si/Mg)$^*$, (C/Mg)$^*$, (O/Mg)$^*$, (Ne/Mg)$^*$, and (He/Mg)$^*$ from 1999 to 2010 obtained by averaging the data points in a given time bin.  For the abundance ratios, the vertical scales are logarithmic, all spanning the same range, 0.8 decades (a factor of 6.3).  For reference, the top panel gives the solar wind alpha particle speed distribution, overlaid with a three-month running average of the mean speed.   Table~\ref{tab_period_stats} provide statistics on the abundance ratios in each of the three time intervals.   The statistical variations in the bulk solar wind parameters, including composition, behave approximately according to a lognormal distribution \citep[see][and references therein]{lepri13}.  We therefore compute the means and standard deviations in lognormal space, and use these values to compute the $1\sigma$ upper and lower bounds of the distributions in linear space that are given in Table~\ref{tab_period_stats}.  These statistics are complemented by Table~\ref{tab_frac_diff} , which tabulates percent differences in the mean abundance values between time periods.

\subsubsection{High-FIP elements relative to Mg}
\label{ssec_highFIP_timevar}

A number of features are readily apparent in the time variation.  The most striking is the large peak in the (C/Mg)$^*$, (O/Mg)$^*$, and (Ne/Mg)$^*$ ratios in 2003.  (Note that this corresponds to a {\it decrease} in fractionation relative to Mg, because the values are approaching unity.)  This clearly tracks with a  large rise in the average monthly solar wind speed, which peaks at about 600 km/s, a good 150 km/s above the typical ecliptic average.  This correspondence is not surprising, being consistent with other observations that the fast (CH) wind is less fractionated than the slow (IS) wind \citep{geiss95, zurbuchen07}.  It also highlights a fortuitous aspect of the Genesis collection period, namely that it included an unexpectedly large period of CH flow, which is unusual for the early declining phase of the solar cycle. 

Aside from the 2003 feature just described, these same three abundance ratios  show a nearly steady rise between the solar maximum and minimum periods.  Referring to Tables~\ref{tab_period_stats} and~\ref{tab_frac_diff}, the bulk SW value for (Ne/Mg)$^*$ increases by 53\% from 0.32 to 0.49; (C/Mg)$^*$ increases by somewhat less, 43\%, from 0.49 to 0.70; and (O/Mg)$^*$ the least, by 30\%, from 0.47 to 0.61.  All of these are real increases, well outside the level of the SWICS instrumental uncertainty.  During the Genesis period, the abundance ratios for these elements have values intermediate between the solar maximum and solar minimum periods, but somewhat closer to the solar maximum values.   The (Ne/Mg)$^*$ ratio shows the steadiest rise with time, whereas (C/Mg)$^*$ and (O/Mg)$^*$ exhibit fluctuations on short time scales.  The most notable example occurs at solar minimum, where (Ne/Mg)$^*$ remains relatively constant but (C/Mg)$^*$ and (O/Mg)$^*$ show a significant dip for an approximately six-month interval in late-2008/early-2009.

Note that the long-period changes in fractionation do not seem to correlate with any obvious long-term trend in the solar wind speed, certainly nothing as strong as the correlation with the speed jump in 2003.  In fact, where the abundances are the least fractionated, in 2003 and then again in 2008-9, the mean SW speed is at its largest in one case and smallest in the other.  This indicates that something other than the relative amounts of IS vs. CH flow are responsible for the observed change in the  apparent FIP fractionation. \citet{lepri13} have also noted changes in elemental composition that were dependent on something other than flow type over the course of the solar cycle. 

We conclude our discussion of the time variation for the High-FIP elements with a few comments on the (He/Mg)$^*$ ratio.  Despite the strong difference in FIP, the (He/Mg)$^*$ stays relatively constant up until the end of 2008, in contrast to the steady climb of the other high-FIP elements.  It is not necessarily surprising that He behaves differently than other high-FIP elements, as the dynamics governing the transport of alpha particles in the corona are unique owing to its large contribution ($\sim 15\%$) to the momentum flux of the solar wind.  What is interesting is that its abundance is so steady compared to magnesium, which seems to indicate they share common conditions for acceleration out of the corona.

The behavior of (He/Mg)$^*$ in late 2008 and thereafter is quite different.  \citet{lepri13} show that in the cycle 23 solar minimum, in the slow solar wind, He/H becomes greatly depleted, and helium even becomes depleted relative to oxygen.   This is consistent with \citet{kasper07} who found that the He/H ratio is strongly correlated with speed below 450 km s$^{-1}$, and essentially goes to zero as the solar wind speed approaches 250~km~s$^{-1}$.  Here, we see that despite the earlier stability of the (He/Mg)$^*$ ratio, there are intervals where helium clearly becomes strongly depleted relative to magnesium, with drops in the   monthly average (He/Mg)$^*$   by over a factor of two.  As we shall show in Section~\ref{sec_hist},  this occurs when the solar wind speed drops below 400~km~s$^{-1}$ In this period, there are also intervals where the (He/Mg)$^*$ monthly average is the highest measured in the solar cycle; thus, the abundance ratio averaged over the whole period is not much changed from the rest of the solar cycle, and is only down by 12\% as compared to solar maximum. 

\subsubsection{Low-FIP elements relative to Mg}
\label{ssec_lowFIP_timevar}

Compared to the high-FIP elements,  (Fe/Mg)$^*$ and (Si/Mg)$^*$ show much less variation over the solar cycle, which is not surprising since Fe, Si and Mg are all part of the same low-FIP class.  However, even for Fe and Si there is visible long-term variation.   Overall, the average (Fe/Mg)$^*$ is constant for the three periods, within the uncertainty limits.  To the eye, (Si/Mg)$^*$ appears slightly but systematically more fractionated during the Genesis period than elsewhere. Compared to solar maximum,  (Si/Mg)$^*$ is 9\% higher in the Genesis period.  That said, this change is within the ~21\% instrumental systematic uncertainty and thus not significant.  
 The  (Fe/Mg)$^*$ ratio shows an anticorrelation with solar wind speed, which is just discernible to the eye (see Fig. 3).  This is consistent with the notion that the fast wind is less fractionated than the slow wind, and it is supported by the regime-specific analysis (see the next subsection).

A more important observation is that the low-FIP elements are consistently fractionated  relative to the photosphere (the Fe/Mg and Si/Mg ratios are significantly above photospheric values) at all times, even though their FIPs are within 0.5 eV of each other.    This discrepancy is more obvious in this study than in most other FIP analyses, as abundances are historically most often normalized to high-FIP ions (e.g.  oxygen or hydrogen) where differences among the low-FIP elements are not as readily discerned.  One would expect that if FIP were the key parameter for determining solar wind elemental abundance, that the low-FIP ratios should be equal to unity to within measurement uncertainty.  Here, we see that these ratios are consistently above unity, by 16-23\% on average for (Fe/Mg)$^*$ and 29-41\% for (Si/Mg)$^*$, depending on the period.  Furthermore, the value of the abundance ratios $1\sigma$ above the mean is $\sim 80\%$ above unity for both (Fe/Mg)$^*$ and (Si/Mg)$^*$.  This is well above the assessed fractionation uncertainty of $\sim 25$\%, a number that combines both the SWICS instrumental uncertainty and the quoted uncertainties of the photospheric abundances of \citet{asplund09} (see also Table~\ref{tab_phot_stats}).   

\subsubsection{Time variation within solar wind regimes}

Tables~\ref{tab_period_stats} and~\ref{tab_frac_diff} also present statistics for the abundance ratios within the three solar wind regimes for each time period.   Here we discuss the differences between the CH, IS and CME  abundances first within a given time period, and then for a given regime across time periods.

Almost without exception, within a given time period, the CH regime was less or equally fractionated as the corresponding IS regime.  The abundance ratios (Fe/Mg)$^*$,  (C/Mg)$^*$, and (O/Mg)$^*$ are all less fractionated  (i.e. closer to 1) in CH than IS flow.  The largest differences were for (C/Mg)$^*$, which showed a 23\%, 34\%, and 33\% change between the IS and CH regimes for the solar maximum, Genesis, and solar minimum periods, respectively. The differences for (O/Mg)$^*$ were about half as much.   And reflecting the discussion in Section~\ref{ssec_lowFIP_timevar}, (Fe/Mg)$^*$ shows small but meaningful differences between IS and CH flow for the solar maximum period (15\%) and the Genesis period (19\%).  

The (Ne/Mg)$^*$ and (He/Mg)$^*$ ratios show no difference of significance between the IS and CH regimes, with the exception of  (He/Mg)$^*$ in the solar minimum period (a 42\% difference).   In the case of (Ne/Mg)$^*$, this is rather curious, since the change in {\it overall} fractionation across the three periods was greater for (Ne/Mg)$^*$ than for any other abundance ratio, as discussed above. The (Si/Mg)$^*$ variations in IS and CH  are within the instrumental uncertainty, therefore the data are consistent within no change in the fractionation.

Of the three sample periods, the Genesis period showed the greatest difference between CH and IS abundances.  This is likely due to the large solar wind speed peak in 2003 and the corresponding shifts in the composition distributions discussed previously in section~\ref{ssec_highFIP_timevar}.

For most species, the CME regime has a composition measurably different from the other two regimes for all time periods. (Although the solar minimum regime shows the strongest difference for CMEs, we do not place much weight on this because there are only seven days of accumulated integration for the CME regime during solar minimum.)   We see that for the (C/Mg)$^*$ and (O/Mg)$^*$ the CME regime is consistently more fractionated than the CH and IS regimes.  This is in keeping with the findings of others for Mg/O and Fe/O \citep{lepri01, richardson04}.  Even the Low-FIP (Si/Mg)$^*$ ratio shows consistent differences between the CME regime and the others.  Interestingly, for (Ne/Mg)$^*$ and (He/Mg)$^*$, as with the CH and IS regimes, the CME regime fractionation is no different.

Regarding the variation of a given regime across time periods: from inspection of Table~\ref{tab_frac_diff} we see that for the most part, the regimes show differences between time periods similar to what was observed for the overall solar wind.  The exception is (He/Mg)$^*$, for which the IS regime is ~30\% more fractionated during solar minimum than the other two periods.  This is a reflection of the extremely low solar wind speeds observed during this solar minimum, and the correspondingly low helium abundances \citep{kasper07}.

\subsubsection{The FIP bias over time}

Here, we define a total FIP bias as the ratio of low-FIP to intermediate and high-FIP elements of all heavy ions. Helium was not included in this definition because it is not a minor ion, and it is of dynamic significance during the solar wind acceleration and expansion process. 
$$f_{FIP} = \frac{[Fe_{sw}+Mg_{sw}+Si{sw}]}{[C_{sw}+O_{sw}+Ne_{sw}]} / \frac{[Fe_{phot}+Mg_{phot}+Si_{phot}]}{[C_{phot}+O_{phot}+Ne_{phot}]}.$$
Values of $f_{FIP}$ for different time periods and regimes are shown in Table 2. For the overall (Bulk) sample, $f_{FIP}$ is lower (by $\sim 7\%$, in the Genesis period compared to solar maximum.  Essentially all of this change occurs for the CH regime, which drops by 15\% (from $f_{FIP} = 2.33$ to $f_{FIP} = 2.03$).   Between the Genesis period and solar minimum, the FIP bias drops again, this time much more significantly, by 25\%.  In this case, the fractional drop is the same for both the CH and IS regimes, although the CH regime (at $f_{FIP} = 1.63$) is about 20\% less fractionated than the IS regime (at $f_{FIP} = 2.12$).

It is interesting to note that of the three periods, solar minimum is the least fractionated overall, and for a given regime. 

\section{Variations of fractionation with solar wind speed}
\label{sec_hist}

Based on the observations of temporal variation that we have set forth here, it is clear that the elemental fractionation of the solar wind varies with FIP, solar wind speed, and possibly other as yet unknown factors.   To further characterize the dependencies of elemental fractionation, in this section we will look in detail at the correlation between fractionation and the solar wind speed.  So far, we have only divided SW measurements into two "speed bins", the CH and IS flow types.  This was  done so that we could compare the Genesis regime samples to what they may have looked like were Genesis flown at other phases of the solar cycle, and also because separation into two flow types (CH/fast and IS/slow) has been historically a standard means of organizing composition.  Such a bifurcation, however, can mask the fact that, say, what is called CH flow during one part of the solar cycle can be quite different than during another part because the speed distributions can be radically different.   It can be seen from the statistics given in Table~\ref{tab_period_stats} for the solar wind speed that the speed distributions within a given regime varied significantly between time periods.  For example, the solar wind speed in the CH regime was $562 \pm 86$ km s$^{-1}$ during the solar maximum period, and $598 \pm 105$ km s$^{-1}$ during the Genesis collection period.  This brings up the possibility that some of the fractionation differences seen in a given regime type but at different phases of the solar cycle may be explained by the differences in the speed distributions within that regime. 

To explore this possibility, we quantify SW fractionation as a function of SW speed within the regimes. Figure~\ref{fig_histograms} shows histogram plots of the lognormal mean fractionation ratios vs SW speed, binned into intervals 50 km s$^{-1}$ wide.  The error bars represent the lognormal standard deviations. For the solar maximum and the Genesis period panels, points are shown specifically for the three different regimes (CH, IS and CME).  For the solar minimum panels, only the CH and IS regimes are represented, because there were too few CMEs to provide a statistically meaningful sample.  For a given regime and speed bin, a point is plotted only if there were at least 30 two-hour time integration measurements available.  All abundances are shown as log values.

The top row of Figure~\ref{fig_histograms} shows the distribution of speed measurements within the different regimes.  Important features to point out are (a) the Genesis collection period shows a substantial number of speed measurements up to 750~km~s$^{-1}$, while the other two periods hardly have any samples above 650~km~s$^{-1}$, and (b) the solar minimum period speed distribution is more skewed toward the lowest speeds than the other two periods, with a sizable number of speed measurements down to 250~km~s$^{-1}$.

\subsection{The speed dependence of the elemental abundance ratios: Evidence of mass-dependent fractionation}
\label{ssec_speed_dep}

Turning to the abundance ratio panels  on Figure 3, the most striking feature is the remarkably linear behavior of $\log(X^*)$ (or exponential behavior in linear space) as a function of speed, particularly for the CH and IS regimes.  In many cases, there is a visible break between the IS and CH regimes, with each following a distinct linear trend.   Linear regression fits have been made to the means within each regime, and are shown by the color-coded lines.  The numerical values for the slopes of the fit lines are given in each panel, in units of $10^{-4} \log(X^*)/$({\rm km~s}$^{-1})$.  Note in the case of the IS or CH regimes, the points that fall in the regions where the speed bins overlap (due to the hysteresis of the regime selection algorithm (see Section~\ref{sec_reg_select} )) tend to be aligned with the points outside the overlap regions.  Two particularly good examples of this can be seen in the (C/Mg)$^*$ and (O/Mg)$^*$ plots for solar maximum.   

The values of log(X$^*$) for the CME regime also vary with speed in a linear fashion, but not quite as regularly as abundance ratios in the other two regimes.  Also, there is a clear and consistent difference in the mean values for the CME regime as compared to the others, and particularly in the case of (C/Mg)$^*$ and (O/Mg)$^*$, the CME regime values are substantially lower, by  $\sim$25\%.

\begin{figure*}[h]
 \centering
 \noindent\includegraphics[trim = 2cm 0cm 0cm 0cm, width=1.1\textwidth]{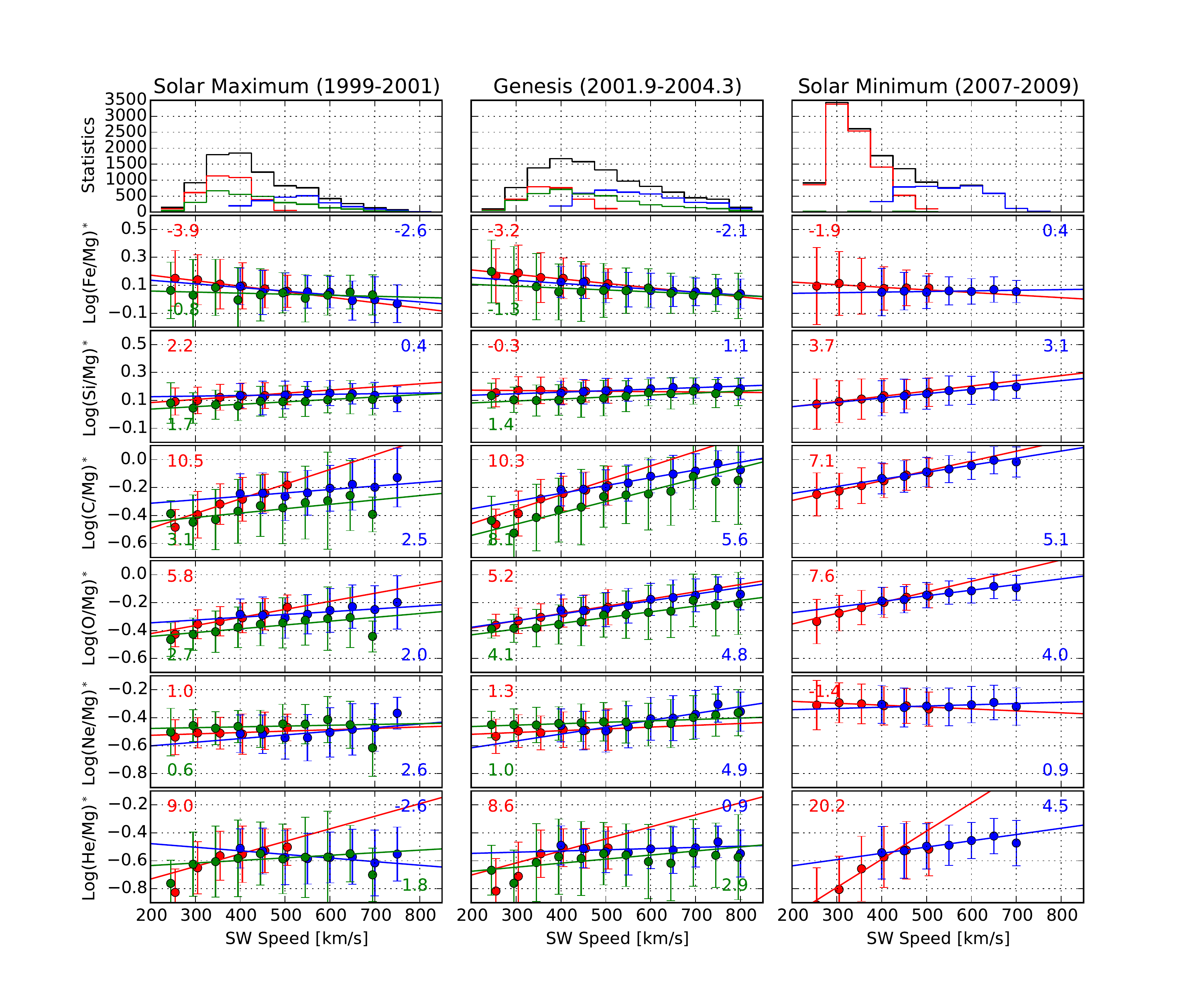}
  \caption{Average fractionation ratios (log scale) binned by velocity and solar wind flow type: interstream (red), coronal hole (blue) and CMEs (green). Data have been binned into 50 km/s intervals. The number of 2-hour observations per bin are shown in the upper panel. The error bars represent standard deviations of the distribution of observations. The linear fits to each regime are plotted, and the numerical values of the slopes of the fits are given.}
\label{fig_histograms}
 \end{figure*}

Organizing the abundance ratio data this way shows that many of the mean abundance differences between regimes (for the same time period) that appeared to be only marginally significant in Table~\ref{tab_period_stats} are in fact real. For example, at solar maximum, the difference between the IS and CH mean abundance for (Fe/Mg)$^*$ is only 15\%, which is just marginally significant, but in Figure~\ref{fig_histograms} we see that (Fe/Mg)$^*$  smoothly drops by 35\% between 250~km~s$^{-1}$ and 800~km~s$^{-1}$; thus, without a doubt these two low-FIP elements are fractionated differently, despite being separated in FIP by only 0.22~eV.   

In fact overall, the variation in fractionation with speed (as quantified by the slopes given in each panel of Figure~\ref{fig_histograms}) is not ordered by FIP at all.  For the high-FIP elements, the largest slopes are for the (C/Mg)$^*$ ratio, and get smaller with increasing FIP.  It does appear, however, that the slopes follow a mass dependent trend: with the exception of (He/Mg)$^*$, the slopes tends to decrease with increasing mass, going from most positive for (C/Mg)$^*$ to most negative for (Fe/Mg)$^*$.  Since we are normalizing to magnesium, one would expect the to see the smallest magnitude slopes for the species closest in mass to Magnesium, which in fact is what we see:  on average the smallest magnitude slopes are seen for (Si/Mg)$^*$ and (Ne/Mg)$^*$.  We note that the fractionation theory of \citet{laming04} based on the action of the pondermotive force predicts a secondary dependence on mass that operates in the same sense as observed here, namely that after FIP fractionation, higher mass elements will experience slightly higher fractionation (\textit{i. e.} an enhanced relative abundance).  The findings presented here suggest that this secondary mechanism may be more effective in the source region of the slow wind.

\subsection{Correcting for differences in speed distributions}

At first glance, it seems quite possible that a large portion of the abundance differences between time periods may be attributable to differences in the speed distributions.  For most ratios,  even the low-FIP (Fe/Mg)$^*$, there is sizable variation with speed.  Thus, if the speed distributions for a given regime are different, then this will naturally give rise to a difference in abundance.  To eliminate the effect of differences in the speed distributions, we have constructed a relationship that compares the abundance ratios between time periods one speed bin at a time.  Specifically, we compute the average percent difference in the speed-specific abundance ratio $X^*$ between time periods:
$$\Delta X^{reg}_{per1-per2} =\frac{1}{n} \sum \limits_{i=1}^n\left( \frac{|X^*_{per1} - X^*_{per2}|}{X^*_{per1}}\right)^{reg}_i \times 100\%,$$
\noindent
where $reg$ is the regime (Bulk, IS, CH or CME); $per1$ and $per2$ are the two time periods being compared; the sum is over $n$ speed bins, and the quantity in parentheses represents the fractional abundance difference between the periods for the {\it i}-th speed bin.   The numerator indicates the absolute value of the difference is to be taken, which is justified because we are interested in differences regardless of sign (an alternative would be to calculate a root-mean-square difference, but this would exaggerate larger differences over smaller ones). Taking the absolute value assures that differences will not be masked by fortuitous cancellation.  Of course, use of the absolute value assures there will always be some difference owing to the presence of statistical fluctuations and systematic error.  However, as with the differences reported in Table~\ref{tab_period_stats}, we do not consider the differences significant unless they are greater than $\leq 21\%$ (15\% for O/Mg).  

We can interpret  the quantity $\Delta X$ as a measure of the difference in the abundances between time periods, corrected for the variation of the speed distribution.  The calculated values of $\Delta X$ are given in Table~\ref{tab_frac_diff_speed}.  The quantities $\Delta X$  can be compared directly to the corresponding entries in Table~\ref{tab_frac_diff}.  If a given $\Delta X$ is significantly less than its counterpart in Table~\ref{tab_frac_diff}, then this would mean that some portion of the abundance differences between time periods is in fact attributable simply to differences in the speed distributions between the periods.  As it turns out, this is hardly ever the case.  In most instances $\Delta X$ is comparable to the corresponding difference in Table~\ref{tab_frac_diff}, and in a few cases, significantly greater.  For example, 
$\Delta ({\rm N/Mg})^{Bulk}_{Gen-Min} = 39\%$, which is comparable to the 40\% difference in the mean values of (Ne/Mg)$^*$ between the Genesis and solar minimum periods.  The case of (C/Mg)$^*$ is one where the abundance differences corrected for differences in the speed distribution are even larger:  $\Delta ({\rm C/Mg})^{IS}_{Gen-Min} = 35\%$, compared to the 21\% difference of the mean values of (C/Mg)$^*$. 

In the cases where all differences are $\leq 21\%$ (15\% for O/Mg),  we can say they are not meaningful, and for all intensive purposes the abundance ratio in question does not vary between time periods.  This looks to be the case for all element comparisons between the Genesis and solar maximum periods, and for the low-FIP element ratios, (Fe/Mg)$^*$ and (Si/Mg)$^*$ across all periods.   For the high-FIP elements, abundance differences between solar minimum and either of the other two periods appear to be caused by changes in the physical conditions in the corona, and not differences in the speed distributions.

There is one case where $\Delta X$ is significantly less than the simple difference of the averages, and that is for (He/Mg)$^*$:  For the IS flow, $\Delta ({\rm He/Mg})^{IS}_{Gen-Min} = 13\%$, below the level of significance, and well below the value of 32\% for the average difference (the same is true comparing the solar maximum and minimum periods).   This is because the primary reason for the difference in the average IS (He/Mg)$^*$ ratios between solar minimum and other times is the shift of the speed distribution to very low speeds, as described in section~\ref{ssec_highFIP_timevar}.

\section{Conclusion}
In this paper we have attempted to characterize the dependencies of solar wind elemental fractionation on conditions in the solar wind with the intention of providing additional useful constraints on models of coronal heating and the mechanism for fractionation.  The recently reanalyzed ACE/SWICS dataset has provided significantly more accurate abundances for the lesser heavy ions in the solar wind, allowing us to reveal new trends.  In addition, by normalizing abundances to the low-FIP ion magnesium, we have uncovered correlations that are not apparent when normalizing to high-FIP ions such as oxygen or hydrogen.

Our principal findings can be summarized as follows:

\begin{enumerate}

\item By normalizing to magnesium, it becomes clear that even the low-FIP elements are measurably fractionated with respect to one another.

\begin{enumerate}
\item When elemental fractionation ratios are sorted more finely by solar wind speed (rather than a simple bifurcation into two categories of  CH/fast wind and IS/slow wind), a clear monotonic trend is observed.  For the IS and CH regimes taken separately, in many cases there is a linear trend in lognormal space (i.e., an exponential dependence on  speed).
 
\item In all cases except (Si/Mg)$^*$, the solar wind becomes less fractionated relative to magnesium with increasing speed.  For (Si/Mg)$^*$, there is no  significant speed dependence.

\item Using our limited set of measured abundances, the fractionation as a function of solar wind speed appears mass-dependent.  Whereas overall fractionation follows a FIP dependence, heavier elements experience a higher degree of fractionation (more enhanced abundance) as the solar wind speed decreases. 

\end{enumerate}
\item For high-FIP elements relative to magnesium, when the fractionation as a function of speed is accounted for, there is still a significant solar cycle dependence to the fractionation.  As suggested by \citet{lepri13}, this could be due to changes in the density of the corona or the amount of wave energy present.   Curiously, this variation is greatest for (Ne/Mg)$^*$  and C/Mg)$^*$, less so for (O/Mg)$^*$  and even less for (He/Mg)$^*$, showing no organization by FIP or mass  for high-FIP elements. 

\item For low-FIP elements, there is no solar cycle dependence to the fractionation, within instrument uncertainties.

\end{enumerate}

The implication for interpretation of the Genesis sample analysis results is that correction of the Genesis-derived elemental abundances to photospheric values is nontrivial.  It will require a model that can reproduce the observed mass-dependent variation in fractionation as a function of solar wind speed, and the variation in abundance over the solar cycle.   For elements observed by SWICS, empirical corrections can be used; however, for other elements analyzed by Genesis, a theoretical model will need to be derived. It is our hope that the findings presented here will help bring about the refinement of such a model in the near future. 

\acknowledgments
We thank the anonymous referee for useful comments. 
P.P., D.B.R. and R.C.W. are supported by NASA Laboratory Analysis of Returned Samples (LARS) program, grants number NNH10A046I and NNX15AG19G. T.H.Z., S.T.L, P.S, J.A.G. and R von S. acknowledge financial support from NASA on grant number  NNX13AH66G.  The authors gratefully acknowledge support from the International Space Science Institute (ISSI).

\bibliographystyle{apj}                       
\bibliography{apj-jour,swpaper.bib}

\begin{thebibliography}{}
\expandafter\ifx\csname natexlab\endcsname\relax\def\natexlab#1{#1}\fi

\bibitem[{{Asplund} {et~al.}(2009){Asplund}, {Grevesse}, {Sauval}, \&
  {Scott}}]{asplund09}
{Asplund}, M., {Grevesse}, N., {Sauval}, A.~J., \& {Scott}, P. 2009, \araa, 47,
  481

\bibitem[{{Bodmer} \& {Bochsler}(1998)}]{bodmer98}
{Bodmer}, R., \& {Bochsler}, P. 1998, \aap, 337, 921

\bibitem[{{Bodmer} \& {Bochsler}(2000)}]{bodmer00}
---. 2000, \jgr, 105, 47

\bibitem[{{Buergi}(1992)}]{buergi92}
{Buergi}, A. 1992, \jgr, 97, 3137

\bibitem[{{Buergi} \& {Geiss}(1986)}]{buergi86}
{Buergi}, A., \& {Geiss}, J. 1986, \solphys, 103, 347

\bibitem[{{Burnett} {et~al.}(2003){Burnett}, {Barraclough}, {Bennett},
  {Neugebauer}, {Oldham}, {Sasaki}, {Sevilla}, {Smith}, {Stansbery},
  {Sweetnam}, \& {Wiens}}]{burnett03}
{Burnett}, D.~S., {Barraclough}, B.~L., {Bennett}, R., {et~al.} 2003, \ssr,
  105, 509

\bibitem[{{Geiss} {et~al.}(1995){Geiss}, {Gloeckler}, \& {von
  Steiger}}]{geiss95}
{Geiss}, J., {Gloeckler}, G., \& {von Steiger}, R. 1995, \ssr, 72, 49

\bibitem[{{Gloeckler} \& {Geiss}(2007)}]{gloeckler07}
{Gloeckler}, G., \& {Geiss}, J. 2007, \ssr, 130, 139

\bibitem[{{Gloeckler} {et~al.}(2003){Gloeckler}, {Zurbuchen}, \&
  {Geiss}}]{gloeckler03}
{Gloeckler}, G., {Zurbuchen}, T.~H., \& {Geiss}, J. 2003, Journal of
  Geophysical Research (Space Physics), 108, 1158

\bibitem[{{Gloeckler} {et~al.}(1998){Gloeckler}, {Cain}, {Ipavich}, {Tums},
  {Bedini}, {Fisk}, {Zurbuchen}, {Bochsler}, {Fischer}, {Wimmer-Schweingruber},
  {Geiss}, \& {Kallenbach}}]{gloeckler98}
{Gloeckler}, G., {Cain}, J., {Ipavich}, F.~M., {et~al.} 1998, \ssr, 86, 497

\bibitem[{{Heber} {et~al.}(2012){Heber}, {Baur}, {Bochsler}, {McKeegan},
  {Neugebauer}, {Reisenfeld}, {Wieler}, \& {Wiens}}]{heber12}
{Heber}, V.~S., {Baur}, H., {Bochsler}, P., {et~al.} 2012, \apj, 759, 121

\bibitem[{{Heber} {et~al.}(2014){Heber}, {McKeegan}, {Smith}, {Jurewicz},
  {Olinger}, {Burnett}, \& {Guan}}]{heber14}
{Heber}, V.~S., {McKeegan}, K.~D., {Smith}, S., {et~al.} 2014, in Lunar and
  Planetary Science Conference, Vol.~45, Lunar and Planetary Science
  Conference, 1203

\bibitem[{{Heber} {et~al.}(2009){Heber}, {Wieler}, {Baur}, {Olinger},
  {Friedmann}, \& {Burnett}}]{heber09}
{Heber}, V.~S., {Wieler}, R., {Baur}, H., {et~al.} 2009, \gca, 73, 7414

\bibitem[{{Kallenbach} {et~al.}(1997){Kallenbach}, {Ipavich}, {Bochsler},
  {Hefti}, {Hovestadt}, {Gr{\"u}nwaldt}, {Hilchenbach}, {Axford}, {Balsiger},
  {B{\"u}rgi}, {Coplan}, {Galvin}, {Geiss}, {Gliem}, {Gloeckler}, {Hsieh},
  {Klecker}, {Lee}, {Livi}, {Managadze}, {Marsch}, {M{\"o}bius}, {Neugebauer},
  {Reiche}, {Scholer}, {Verigin}, {Wilken}, \& {Wurz}}]{kallenbach97}
{Kallenbach}, R., {Ipavich}, F.~M., {Bochsler}, P., {et~al.} 1997, \jgr, 102,
  26895

\bibitem[{{Kallenbach} {et~al.}(1998){Kallenbach}, {Ipavich}, {Kucharek},
  {Bochsler}, {Galvin}, {Geiss}, {Gliem}, {Gloeckler}, {Gr{\"u}nwaldt},
  {Hefti}, {Hilchenbach}, \& {Hovestadt}}]{kallenbach98}
{Kallenbach}, R., {Ipavich}, F.~M., {Kucharek}, H., {et~al.} 1998, \ssr, 85,
  357

\bibitem[{{Kasper} {et~al.}(2007){Kasper}, {Stevens}, {Lazarus}, {Steinberg},
  \& {Ogilvie}}]{kasper07}
{Kasper}, J.~C., {Stevens}, M.~L., {Lazarus}, A.~J., {Steinberg}, J.~T., \&
  {Ogilvie}, K.~W. 2007, \apj, 660, 901

\bibitem[{{Laming}(2004)}]{laming04}
{Laming}, J.~M. 2004, \apj, 614, 1063

\bibitem[{{Laming}(2009)}]{laming09}
---. 2009, \apj, 695, 954

\bibitem[{{Lepri} {et~al.}(2013){Lepri}, {Landi}, \& {Zurbuchen}}]{lepri13}
{Lepri}, S.~T., {Landi}, E., \& {Zurbuchen}, T.~H. 2013, \apj, 768, 94

\bibitem[{{Lepri} {et~al.}(2001){Lepri}, {Zurbuchen}, {Fisk}, {Richardson},
  {Cane}, \& {Gloeckler}}]{lepri01}
{Lepri}, S.~T., {Zurbuchen}, T.~H., {Fisk}, L.~A., {et~al.} 2001, \jgr, 106,
  29231

\bibitem[{{Marsch} {et~al.}(1995){Marsch}, {von Steiger}, \&
  {Bochsler}}]{marsch95}
{Marsch}, E., {von Steiger}, R., \& {Bochsler}, P. 1995, \aap, 301, 261

\bibitem[{{Meyer}(1985)}]{meyer85}
{Meyer}, J.-P. 1985, \apjs, 57, 173

\bibitem[{{Meyer}(1993)}]{meyer93}
{Meyer}, J.-P. 1993, in Origin and Evolution of the Elements, ed.
  N.~{Prantzos}, E.~{Vangioni-Flam}, \& M.~{Casse}, 26--62

\bibitem[{{Neugebauer} {et~al.}(2003){Neugebauer}, {Steinberg}, {Tokar},
  {Barraclough}, {Dors}, {Wiens}, {Gingerich}, {Luckey}, \&
  {Whiteaker}}]{neugebauer03}
{Neugebauer}, M., {Steinberg}, J.~T., {Tokar}, R.~L., {et~al.} 2003, \ssr, 105,
  661

\bibitem[{{Reisenfeld} {et~al.}(2003){Reisenfeld}, {Steinberg}, {Barraclough},
  {Dors}, {Wiens}, {Neugebauer}, {Reinard}, \& {Zurbuchen}}]{reisenfeld03}
{Reisenfeld}, D.~B., {Steinberg}, J.~T., {Barraclough}, B.~L., {et~al.} 2003,
  in American Institute of Physics Conference Series, Vol. 679, Solar Wind Ten,
  ed. M.~{Velli}, R.~{Bruno}, F.~{Malara}, \& B.~{Bucci}, 632--635

\bibitem[{{Reisenfeld} {et~al.}(2013){Reisenfeld}, {Wiens}, {Barraclough},
  {Steinberg}, {Neugebauer}, {Raines}, \& {Zurbuchen}}]{reisenfeld13}
{Reisenfeld}, D.~B., {Wiens}, R.~C., {Barraclough}, B.~L., {et~al.} 2013, \ssr,
  175, 125

\bibitem[{{Reisenfeld} {et~al.}(2007){Reisenfeld}, {Burnett}, {Becker},
  {Grimberg}, {Heber}, {Hohenberg}, {Jurewicz}, {Meshik}, {Pepin}, {Raines},
  {Schlutter}, {Wieler}, {Wiens}, \& {Zurbuchen}}]{reisenfeld07}
{Reisenfeld}, D.~B., {Burnett}, D.~S., {Becker}, R.~H., {et~al.} 2007, \ssr,
  130, 79

\bibitem[{{Richardson} \& {Cane}(2004)}]{richardson04}
{Richardson}, I.~G., \& {Cane}, H.~V. 2004, Journal of Geophysical Research
  (Space Physics), 109, 9104

\bibitem[{{Richardson} \& {Cane}(2010)}]{richardson10}
---. 2010, \solphys, 264, 189

\bibitem[{{Schwadron} {et~al.}(1999){Schwadron}, {Fisk}, \&
  {Zurbuchen}}]{schwadron99}
{Schwadron}, N.~A., {Fisk}, L.~A., \& {Zurbuchen}, T.~H. 1999, \apj, 521, 859

\bibitem[{{Shearer} {et~al.}(2014){Shearer}, {von Steiger}, {Raines}, {Lepri},
  {Thomas}, {Gilbert}, {Landi}, \& {Zurbuchen}}]{shearer14}
{Shearer}, P., {von Steiger}, R., {Raines}, J.~M., {et~al.} 2014, \apj, 789, 60

\bibitem[{{Stone} {et~al.}(1998){Stone}, {Frandsen}, {Mewaldt}, {Christian},
  {Margolies}, {Ormes}, \& {Snow}}]{stone98}
{Stone}, E.~C., {Frandsen}, A.~M., {Mewaldt}, R.~A., {et~al.} 1998, \ssr, 86, 1

\bibitem[{{Tu} \& {Marsch}(1995)}]{tu95}
{Tu}, C.-Y., \& {Marsch}, E. 1995, \ssr, 73, 1

\bibitem[{{von Steiger} \& {Zurbuchen}(2011)}]{vonsteiger11}
{von Steiger}, R., \& {Zurbuchen}, T.~H. 2011, Journal of Geophysical Research
  (Space Physics), 116, 1105

\bibitem[{{von Steiger} {et~al.}(2000){von Steiger}, {Schwadron}, {Fisk},
  {Geiss}, {Gloeckler}, {Hefti}, {Wilken}, {Wimmer-Schweingruber}, \&
  {Zurbuchen}}]{vonsteiger00}
{von Steiger}, R., {Schwadron}, N.~A., {Fisk}, L.~A., {et~al.} 2000, \jgr, 105,
  27217

\bibitem[{{Zhao} {et~al.}(2009){Zhao}, {Zurbuchen}, \& {Fisk}}]{zhao09}
{Zhao}, L., {Zurbuchen}, T.~H., \& {Fisk}, L.~A. 2009, \grl, 36, 14104

\bibitem[{{Zurbuchen}(2007)}]{zurbuchen07}
{Zurbuchen}, T.~H. 2007, \araa, 45, 297

\bibitem[{{Zurbuchen} {et~al.}(2002){Zurbuchen}, {Fisk}, {Gloeckler}, \& {von
  Steiger}}]{zurbuchen02}
{Zurbuchen}, T.~H., {Fisk}, L.~A., {Gloeckler}, G., \& {von Steiger}, R. 2002,
  \grl, 29, 1352

\end{thebibliography}
\end{document}